# Conformal frequency conversion for arbitrary vectorial structured light

Hai-Jun Wu,[1] Bing-Shi Yu,[1] Zhi-Han Zhu,[1,*] Wei Gao,[1] Dong-Sheng Ding,[1,2] Zhi-Yuan Zhou,[1,2] Xiao-Peng Hu,[3] Carmelo Rosales-Guzmán,[1,4] Yijie Shen[5] and Bao-Sen Shi[1,2,†]

[1] *Wang Da-Heng Center, Heilongjiang Key Laboratory of Quantum Control, Harbin University of Science and Technology, Harbin 150080, China*
[2] *CAS Key Laboratory of Quantum Information, University of Science and Technology of China, Hefei, 230026, China*
[3] *National Laboratory of Solid State Microstructures, College of Engineering and Applied Sciences, Nanjing University, Nanjing 210093, China*
[4] *Centro de Investigaciones en Óptica, A.C., Loma del Bosque 115, Colonia Lomas del campestre, 37150 León, Gto., Mexico*
[5] *Optoelectronics Research Centre, University of Southampton, Southampton SO17 1BJ, United Kingdom*

Vectorial structured light with spatially varying amplitude, phase, and polarization is reshaping many areas of modern optics, including nonlinear optics, as diverse parametric processes can be used to explore interactions between such complex vector fields, extending the frontiers of optics to new physical phenomena. However, the most basic nonlinear application, i.e., frequency conversion, still remains challenging for vectorial structured light since parametric processes are polarization dependent, leading to a change in the spatial topological structure of signals. In this work, to break this fundamental limit, we propose a novel conformal frequency conversion scheme that allows to maintain the full spatial structure of vectorial structured light in the conversion; and systematically examine its spatial polarization independence based on non-degenerate sum-frequency generation with type-0 phase matching. This proof-of-principle demonstration paves the way for a wide range of applications requiring conformal frequency conversion, and, particularly, to implement frequency interfaces with multimodal communication channels, high-dimensional quantum states, and polarization-resolved upconversion imaging.

## I. Introduction

The frequency conversion of light waves by means of diverse parametric processes is one of the core applications in nonlinear optics because it provides an additional way to control the longitudinal mode (i.e., frequency distribution) of beams generated by laser cavities [1,2]. This critical process forms the physical foundation for realizing a wide variety of modern photonic techniques, such as the parametric laser, upconversion/downconversion detection, and the frequency interface for optical signals [3–5]. Among them, the interactions of second-order optical parameters, including sum- and down-frequency generations, are the most used nonlinear processes. It is also wort mentioning there are various types of quasi-phase matchings (QPMs), proposed as early as 1962 by Bloembergen [6], that after decades of advances in microstructure fabrication have been realized in various domain-engineered ferroelectric crystals [7]. This holds especially well for the so-called type-0 QPM ($e + e \rightarrow e$ or $o + o \rightarrow o$), in which the largest nonlinear coefficient $d_{33}$ in a long crystal can be exploited without inducing the walk-off effect. In this way, the intensity of light–matter interactions can be boosted several orders of magnitude. Based on the high-efficient QPM, now the frequency conversion can be accessed to applications with weak light and even to single photons [8,9].

In addition to the longitudinal mode, light has a transverse structure involving amplitude, phase, and polarization (or spin). For a paraxial field, the structure can be described fully and economically by its state of polarization (SoP) and spatial mode, or, in some cases, by a non-separable superposition of the two that is known as vectorially spatial mode (vector mode for short) [10–13]. Owing to recent progress in the spatial modulation of light [14,15], the field of structured light, which focuses on the generation, reshaping, and application of the transverse structure of paraxial fields, has completely reshaped the landscape of modern optics from fundamental physics to advanced photonic techniques [16–23]. This is the case of nonlinear optics, which has also greatly benefited from the study of structured light. The corresponding studies in the area began by exploring the conservation of the orbital angular momentum (OAM) in various optical parametric interactions [24–26], which subsequently attracted significant interest in the context of high-dimensional quantum information [27–29]. More recently, the research has focused on ways to shape the full-field transverse structure of light via nonlinear transformations to exploit all dimensions and degrees of freedom [30–35].

* zhuzhihan@hrbust.edu.cn
† drshi@ustc.edu.cn

The frequency conversion of vector modes is perhaps a critical step forward to finally open nonlinear optics to a new world of modern applications—for instance, it can provide a frequency interface for multimodal communication channels or high-dimensional quantum states [36–38]. However, this remains challenging because parametric interactions are heavily dependent on both SoP and amplitude, whereas vector modes have spatially varying SoPs and amplitudes that originate from photonic spin–orbit coupling (SOC). As a result, the spatial amplitude of the excited nonlinear polarization (i.e., dipole oscillation in the media) becomes a function of both the applied vector mode and the polarization relation of the phase matching condition [39–42]. Thus, a conformal frequency convertor for vector modes should be spatial polarization independent; in other words, it should be independent of the SoPs and spatial modes of the input signals. The most feasible way to achieve this is by using a Sagnac nonlinear interferometer [43], which was originally proposed for spontaneous down-conversion by Shi and Tomita [44]. It has since become the most popular source of entanglement used in labs and even on satellites owing to its high efficiency and phase stability [45,46]. Another well-known source of entanglement, i.e., a two-crystal sandwich that is simple to construct, can be used for the parametric conversion of vector modes [47,48], but it is not compatible with long QPM crystals and thus lacking efficiency.

In a previous work, we experimentally demonstrated the frequency-doubling of beams with cylindrical vector (CV) polarizations by using a polarization interferometer with a type-II QPM crystal [49], which preceded the report of a similar experiment [50]. Crucially, the type-II QPM is used to generate polarization-entangled photons, but cannot offer large nonlinear coefficient. Its efficiency for frequency conversion is approximately only 10% of that of the type-0 QPM. In this work, we study the conformal upconversion of arbitrary vector modes via non-degenerate sum-frequency generation (SFG) based on the type-0 QPM. By using the Sagnac nonlinear interferometer with a long type-0 crystal, we show that arbitrary (including both 2D and 3D) vector modes in the near-infrared (telecom) band can be upconverted conformally into the visible band (green) without changing their vectorial spatial structure. The remainder of this paper is organized as follows: In Sec. II, we introduce the theoretical principle for realizing conformal upconversion. Here we also provide details of the experimental implementation of this technique. In Sec. III we provide experimentally results showing the conformal upconversion of arbitrary vector modes to validate our theoretical and experimental methods. The conclusions of this study are given in Sec. IV, and a detailed theoretical description and additional experimental results are provided in Supplementary Materials (SM).

## II. Methods

*Principle of conformal upconversion.* — A general paraxial vector mode can be expressed as a non-separable superposition of a pair of orthogonal SoPs $\hat{e}_\pm$ and associated orthogonal spatial modes $\psi_\pm(\mathbf{r},z)$, given by

$$\mathbf{E}(\mathbf{r},z) = \sqrt{\alpha}\psi_+(\mathbf{r},z)\hat{e}_+ + e^{i\theta}\sqrt{1-\alpha}\psi_-(\mathbf{r},z)\hat{e}_-,$$
$$\psi_\pm(\mathbf{r},z) = u_\pm(\mathbf{r},z)e^{-ik(\omega)z} \quad (1)$$

where $\alpha \in [0,1]$ is a weighting coefficient, $\theta$ is an initial intramodal phase, $k(\omega)$ is the dispersion relation, and $\mathbf{r}$ denotes the transverse coordinates. In particular, $\mathbf{E}(\mathbf{r},z)$ would have a self-similar (i.e., propagation-invariant) vector profile when $\psi_\pm(\mathbf{r},z)$ have the same modal order $N$. Specifically, the pattern of intensity $\alpha u_+^2(\mathbf{r},z) + (1-\alpha)u_-^2(\mathbf{r},z)$ and the associated spatially variant SoP are both invariant upon propagation, except for an overall enlarging by $\sqrt{2}$ per Rayleigh distance ($z_R$) [51]. For a given orthogonal basis pair $\psi_\pm(\mathbf{r},z)\hat{e}_\pm$, all possible vector modes form a hybrid parameter space that obeys SU(2) symmetry, and can be visualized as the surface of an unit sphere. A particular case is the so-called higher-order Poincaré sphere (HOPS) that was introduced to describe paraxial SOC states, i.e., CV modes, in which $\psi_\pm(\mathbf{r},z)$ represents a pair of conjugate Laguerre–Gauss (LG) modes carrying opposite OAM [52].

The conformal frequency convertor for the vector modes shown in Eq. (1) should be able to keep both spatial modes (i.e., $\psi_\pm(\mathbf{r},z)$) and associated parameters (i.e., $\alpha$ and $\theta$) constant; in other words, the position of a vector mode on its unit sphere before and after the conversion should remain constant. To achieve this, we use an SU(2) nonlinear interferometer based on the Sagnac polarization scheme with a long type-0 QPM crystal, as shown in the bottom-right inset of Fig. 1. It works as follows: A signal vector mode with frequency $\omega_1$, given by $\mathbf{E}_S^{\omega_1}(\mathbf{r},z)$, and a diagonally polarized flattop pump (i.e., super-Gauss mode) with frequency $\omega_2$, given by $E_P^{\omega_2}(\mathbf{r},z)\hat{e}_D$, are injected into the Sagnac loop from port-1 and port-2, respectively. The input signal and the pump are first split by a three-wavelength polarizing beam splitter (*t*-PBS), which in the Sagnac loop can be represented as

$$\mathbf{E}_S^{\omega_1}(\mathbf{r},z) = E_H^{\omega_1}(\mathbf{r},z)\hat{e}_H + e^{i\phi}E_V^{\omega_1}(\mathbf{r},z)\hat{e}_V$$
$$E_P^{\omega_2}(\mathbf{r},z)\hat{e}_D = u_P(\mathbf{r},z)e^{-ik(\omega_2)z} \times 1/\sqrt{2}(\hat{e}_H + \hat{e}_V), \quad (2)$$

where $E_{H/V}^{\omega_1}(\mathbf{r},z)$ are SoP-dependent spatial modes with respect to $\hat{e}_H$ and $\hat{e}_V$ that propagate in the clockwise and anticlockwise directions in the loop, respectively, and $e^{i\phi}$ is the associated intramodal phase (see Appendix A for details). Note that, first, $E_{H/V}^{\omega_1}(\mathbf{r},z)$ usually are not orthogonal to one another, except for when the signal is in its maximum non-separable SOC state, i.e., $\alpha = 0.5$. Second, the spatial amplitude of the pump $u_P(\mathbf{r},z)$ should be spatially homogeneous, i.e., a super-Gauss beam, to maintain the high-order spatial mode during upconversion [53]. In the Sagnac loop, a long type-0 QPM ( $\hat{e}_V + \hat{e}_V \rightarrow \hat{e}_V$ ) crystal satisfying $k(\omega_1) + k(\omega_2) = k(\omega_3)$ is used to perform the non-degenerate SFG. The presence of an achromatic-half-wave plate (*a*-HWP) or a Fresnel rhomb prism, set at 45° for swapping $\hat{e}_H$ and $\hat{e}_V$, ensures that the type-0 QPM is simultaneously valid for clockwise and anticlockwise trips. In this way, the nonlinear transformation in the clockwise propagation is $E_H^{\omega_1}\hat{e}_H * E_P^{\omega_2}\hat{e}_H \rightarrow E_H^{\omega_1}\hat{e}_V * E_P^{\omega_2}\hat{e}_V \rightarrow E_H^{\omega_3}\hat{e}_V$ while that in the anticlockwise propagation is $E_V^{\omega_1}\hat{e}_V * E_P^{\omega_2}\hat{e}_V \rightarrow E_V^{\omega_3}\hat{e}_V \rightarrow E_V^{\omega_3}\hat{e}_H$. Moreover, the intramodal phase $e^{i\phi}$ can be adequately maintained by the self-stable structure of the Sagnac loop. As a consequence, the finally upconverted light with frequency $\omega_3$ obtained by port-3 is still in the vector mode of the input signal, which can be expressed as

$$\mathbf{E}_U^{\omega_3}(\mathbf{r}, z) = E_H^{\omega_3}(\mathbf{r}, z)\hat{e}_H + e^{i\phi}E_V^{\omega_3}(\mathbf{r}, z)\hat{e}_V, \quad (3)$$

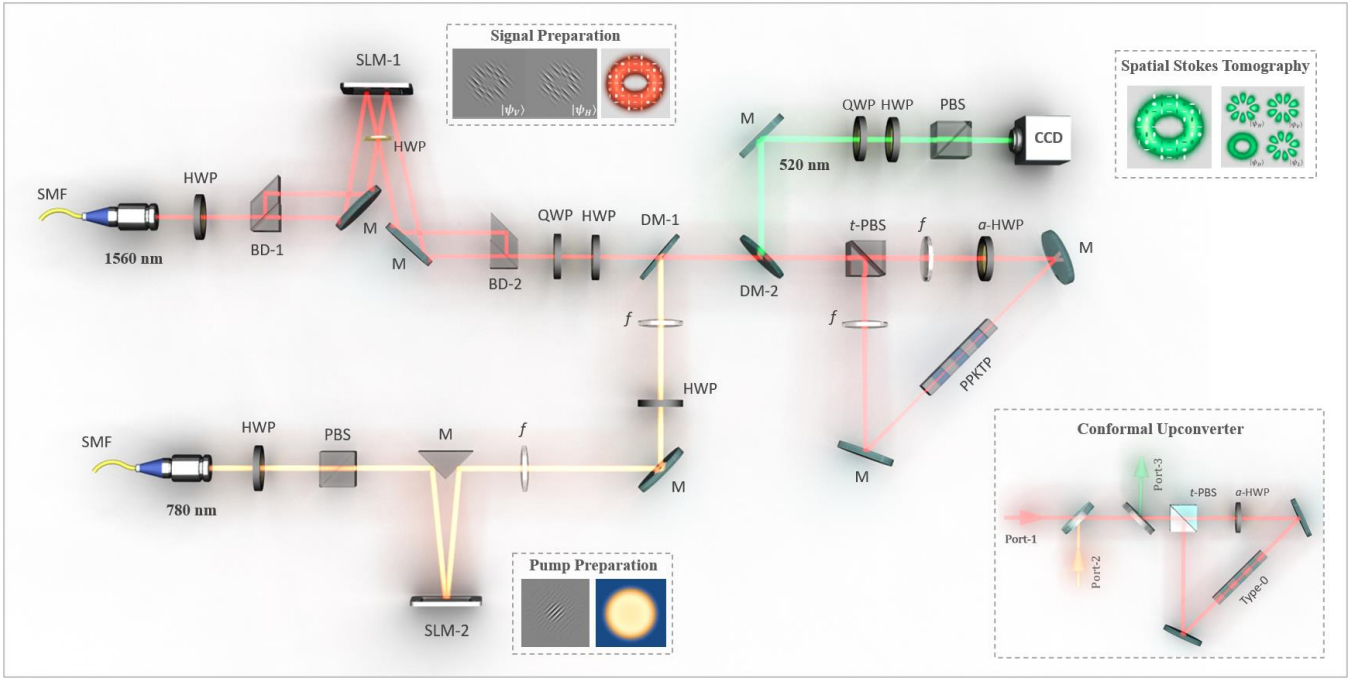

FIG. 1. Diagram of the experimental setup. The key components include a single-mode fiber (SMF), polarizing beam splitter (PBS), half-wave plate (HWP), quarter-wave plate (QWP), polarizing beam displacer prism (BD), mirror (M), lens (*f*), dichroic mirror (DM), spatial light modulator (SLM), *three-wavelength* polarizing beam splitter (*t*-PBS), and *achromatic* half-wave plate (*a*-HWP). The inset in the upper-right corner shows the principle of spatial Stokes measurement while that in the right bottom shows the principle of the Sagnac nonlinear interferometer for conformal upconversion.

In this way, conformal upconversion for arbitrary vector modes is accomplished.

*Experimental setup.* —Figure 1 shows the setup used to verify the principle explained above. A laser with a narrow linewidth (New Focus TLB-6728 with EDFA) operating at 1560 nm and its frequency-doubling (780 nm) were used to generate the signal and the pump beams, respectively. To begin with, both beams were first spatially filtered with a single-mode fiber collimator to obtain a TEM$_{00}$ mode. The pump light was then sent to a spatial light modulator (SLM-1, Holoeye PLUTO-2-080) and converted afterwards into the optimum flattop pump beam (see Ref. 13 for technical details). The prepared pump beam is then relayed to the input port of the conformal upconverter, i.e., the Sagnac loop, by an imaging system. The infrared signal was first sent into a vector mode generator, which was a self-stable polarization Mach–Zehnder interferometer consisting of polarizing beam displacers (BD), a half-wave plate (HWP), and SLM-2 (Holoeye PLUTO-2-013) (see Refs. 54 for details). To generate the desired vector mode, e.g., $\sqrt{\alpha}\psi_H(\mathbf{r}, z)\hat{e}_H + e^{i\theta}\sqrt{1-\alpha}\psi_V(\mathbf{r}, z)\hat{e}_V$, the infrared beam was separated by BD-1 into two parallel beams with orthogonal SoPs and a power ratio of $\alpha/(1-\alpha)$. Their spatial modes were then tailored into $\psi_{H/V}$ by two sections of SLM-2 via complex amplitude modulation [14]. After recombining both beams at BD-2, a group of waveplates were used to further convert the SoP basis into the desired one.

The prepared signal and pump beams were combined via a dichroic mirror (DM-1) and then injected into the Sagnac loop via DM-2 (transmitting at 650–1700 nm). In the loop, the co-propagating pump and signal, for both clockwise and anticlockwise trips, were focused on a type-0 periodically poled KTiOPO$_4$ (PPKTP) crystal (1×2×20 mm) by using a pair of lenses with a focal length of 100 mm in the loop. To control and lock the temperature of the QPM, the crystal was placed in a temperature-controlled oven with a stability of ±2 mK [55]. The upconverted $\psi_{H/V}$ (520 nm) in the two trips were recombined as a vector mode, i.e., $\mathbf{E}_U^{\omega_3}(\mathbf{r},z)$, which was finally exported from port-3 at DM-2 (reflecting at 350–650 nm). Afterwards, the output upconversion was characterized by spatial Stokes tomography, performed through a series of SoP projections and a charge coupled device (CCD) camera [56, 57]. The optical elements of the high-quality polarization ratio are crucial for achieving conformal upconversion with a high fidelity. For this, the extinction ratios of the t-PBS were 1500:1 and 500:1 for the transmission and reflection ports, respectively, whereas the retardances of a-HWP (Thorlabs SAHWP05M-1700) were 0.5, 0.5, and 0.47 waves at 1560, 780, and 520 nm, respectively.

## III. Results

In this section, we corroborate experimentally the principle and methods introduced in Sec. II. To this end, the performance of the system was first characterized. Afterwards, conformal upconversion for various eigen and non-eigen vector modes, including generalized SOC modes, self-imaging (or so-called Talbot) vector modes, and finally optical skyrmions, were then performed. For convenience, we used LG modes, denoted by $LG_p^{\pm \ell}$, and their superpositions to represent general spatial modes, where $\ell$ and $p$ are the azimuthal and radial indices, respectively (see Appendix A in SM for details).

### A. Characterization of the conformal conversion

We focus on two crucial capabilities of the system, i.e., spatial mode-independent efficiency of conversion and intramodal phase consistency, that are afforded by both the apparatus and the flattop pump.

To focus on the influence of spatial modes on the nonlinear interaction, we considered only small-signals in undepleted regions, where the efficiency of upconversion, in both the power ($\eta_p$) and quantum ($\eta_q$) scenarios, increased linearly with pump power. We also measured the efficiency of conversion in the signal-depleted region by using pulsed beams, as shown in Appendix B of SM. About the two efficiencies, first, the amplitude of the SFG is determined by the amplitude of the nonlinear polarization ($\mathbf{P}^{NL}$), which is proportional to the amplitude of coupling of the mixing waves. We thus have the relation $\eta_p \propto \mathbf{P}^{NL} \propto \kappa E_p^{\omega 1} E_s^{\omega 2}$ that indicates, for a given SFG, that power efficiency ($\eta_p$) is consistent for $\omega_1$ or $\omega_2$ as signal, and could finally exceed one in the region of depletion. Second, quantum efficiency ($\eta_q$) is defined by how many photons are converted, and thus further depends on the frequencies (i.e., energy) of the signal photons, given by $\eta_q = \eta_p \times (\omega_1/\omega_3)$. In the experiments, the powers of the pump (780 nm) and the signal (1560 nm) were set to 1 W and 1 mW, respectively. The original waists of the 780 and 1560 nm beams were set to 1 and 2 mm, respectively, so that their fundamental modes, i.e., $LG_0^0$, at the focal plane perfectly overlapped with each other, as shown in the right side of Fig. 2(a).

Figure 2(a) shows the measured quantum efficiencies (%/W pump) for various high-order LG modes in the Gauss-pumped upconversion, i.e., $\eta_q(\ell, p)$. The results show that $\eta_q(\ell, p)$ decreased with both the azimuthal and the radial indices of the signals, which can be interpreted as the intensity of $\mathbf{P}^{NL}$ also depends on the overlap between mixing waves. This mode-dependent efficiency is a critical bottleneck for realizing conformal upconversion, leading to mode distortion in upconversion, which is undesired for the quantum control of high-dimensional photonic states. This becomes even challenger when the signal carries non-zero radial indices, because the modulation of spatial intensity imposed by the pump gives rise to the degeneration of signal's radial components [30], which is a thorny issue even for classical applications (see Appendix A of SM for details). A feasible way to solve this problem is by using a super-Gauss beam as a pump to realize a "flattop-shaped" efficiency of conversion for high-order modes [53]. In our experiments, we used a computed hologram loaded on a SLM to generate a super-Gaussian mode with a customized transverse shape and size [14]. Figure 2(b) shows the measured quantum efficiency of the SFG pumped by the super-Gauss beam, where the pump was optimized to adequately match the size of $LG_2^{\pm 4}$ within the crystal, as shown in the right-side inset. Even though it reduced the efficiency of low-order modes, depending on the maximum modal size of signal, a flattop-shaped efficiency of conversion ranging from $LG_0^0$ to $LG_2^{\pm 4}$ was finally achieved. More importantly, the super-Gauss pump could adequately maintain the spatial mode spectrum of the signal, thus providing a key basis for conformal upconversion.

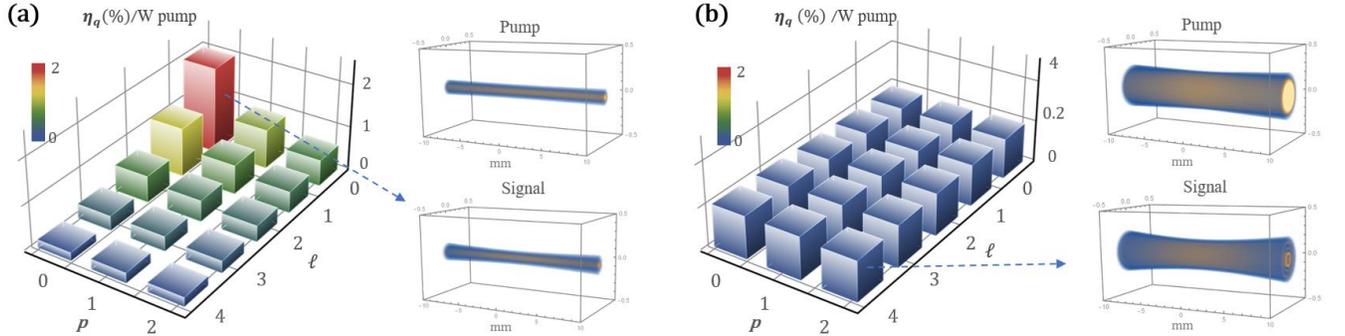

FIG. 2. (a) Measured quantum efficiencies of various high-order LG modes in upconversion pumped by a 1 W Gauss beam, where the right-side schematic indicates that only the $LG_0^0$ signal adequately overlapped with the pump within the crystal. (b) The measured quantum efficiencies of various high-order LG modes in flattop-beam-pumped upconversion, where even $LG_2^4$ overlapped well with the pump.

We now further characterize the performance of our system in maintaining the SoP, i.e., the consistency of $\alpha$ and $\theta$ as passing the SU(2) interferometer. For convenience, we considered mutually unbiased bases (MUBs) in the SOC space of the CV modes, introduced by Milione *et al.* [52], that are given by

$$\text{I} = \left\{ |L_\ell\rangle = |LG_0^{+\ell}, \hat{e}_L\rangle, \; |R_\ell\rangle = |LG_0^{-\ell}, \hat{e}_R\rangle \right\};$$
$$\text{II} = \left\{ |H_\ell\rangle = (|L_\ell\rangle + |R_\ell\rangle)/\sqrt{2}, \; |V_\ell\rangle = (|L_\ell\rangle - |R_\ell\rangle)/\sqrt{2} \right\};$$
$$\text{III} = \left\{ |D_\ell\rangle = (|H_\ell\rangle + |V_\ell\rangle)/\sqrt{2}, \; |A_\ell\rangle = (|H_\ell\rangle - |V_\ell\rangle)/\sqrt{2} \right\}. \quad (4)$$

Figure 3(a) shows the theoretical vector profiles (including the structures of both intensity and polarization) of the MUBs with $\ell = 1$. The yellow points in Fig. 3(d) show their corresponding positions on the HOPS. Owing to their unique transverse structures, these vector modes are important photonic carriers for alignment-free and high-dimensional quantum communication [36, 37, 58–60]. Figures 3(b) and (c) show the vector profiles of the prepared signals (1560 nm) and their corresponding upconversion (520 nm), respectively. They agreed well with each other as well as with their theoretical references, shown in Fig. 3(a). To quantitatively show this excellent agreement, the SOC states of both signals and their upconversion were obtained via spatial Stokes tomography [56, 57]. These SOC states were geometrically represented as red and blue points on a HOPS, as shown in Fig. 3(d), and corresponding correlation matrices of the MUBs are shown in Fig. 3(e). The data showed that all signal beams, prepared according to the theory, were perfectly upconverted. All the above results verify the excellent performance of the experimental platform.

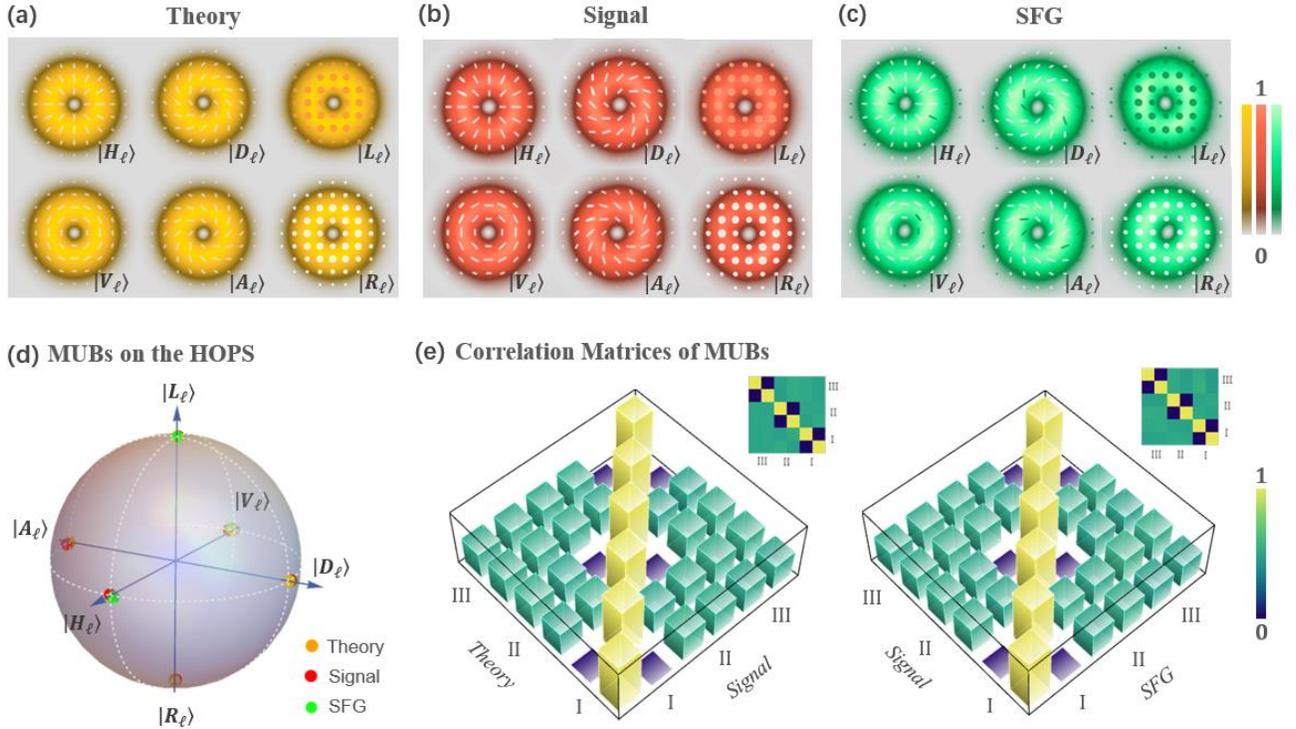

FIG. 3. Conformal upconversion of complete MUBs of the SOC space spanned by $|L\rangle = |\hat{e}_L, +1\rangle$ and $|R\rangle = |\hat{e}_R, -1\rangle$. (a)–(c) show the theoretical vector profiles, the measured signals, and the corresponding SFG, respectively, where the ellipses on the beam profiles depict the spatially variant SoP. (d) Positions of the MUBs on the HOPS. (e) Experimental correlation matrix for different MUBs obtained from the measured signals and the corresponding SFG, respectively.

## B. Conformal upconversion of generalized SOC modes

In this subsection we demonstrate the conformal upconversion of generalized SOC modes. These signals are still vector eigenmodes of the paraxial wave equation, and their transverse structures are thus propagation invariant but, unlike CV modes, they do not have rotational symmetry. In any coordinates and for a given modal order $N$, there are always two orthogonal and complementary eigenmodes that have the same intensity profile but carry the opposite OAM. Namely, they can form a SU(2) OAM space, as well as the associated SOC space. For instance, a pair of conjugate LG modes and associated CV modes are those with respect to cylindrical coordinates. We now show the conformance of our platform for generalized SOC modes in elliptical coordinates, which are the smooth transition between the cylindrical and Cartesian coordinates by changing the ellipticity of the system.

These elliptic SOC modes refer to vectorial Ince–Gauss (IG) modes [61], which can be expressed as

$$\sqrt{\alpha} IG_{Nm}^+(\mathbf{r},z;\varepsilon)\hat{e}_+ + e^{i\theta}\sqrt{1-\alpha} IG_{Nm}^-(\mathbf{r},z;\varepsilon)\hat{e}_-, \quad (5)$$

where $IG_{Nm}^\pm$ denotes the $N$-order helical IG modes with the same intensity but carrying the opposite OAM, and $\varepsilon \in [0,\infty)$ defines the ellipticity of the mode. The relation between the helical IG modes and their counterparts with (even/odd) parity is given by $IG_{Nm}^\pm = \sqrt{1/2}(IG_{Nm}^e \pm iIG_{Nm}^o)$ (see Appendix A of SM for details), and the helical modes usually carry the non-integer OAM as $\varepsilon \neq 0$ [62,63]. Note that the elliptic coordinates are a function of $\varepsilon$; in particular, as $\varepsilon = 0$ and $\varepsilon \to \infty$, the IG modes were converted into the LG and Hermite–Gauss (HG) modes with the same modal orders, respectively.

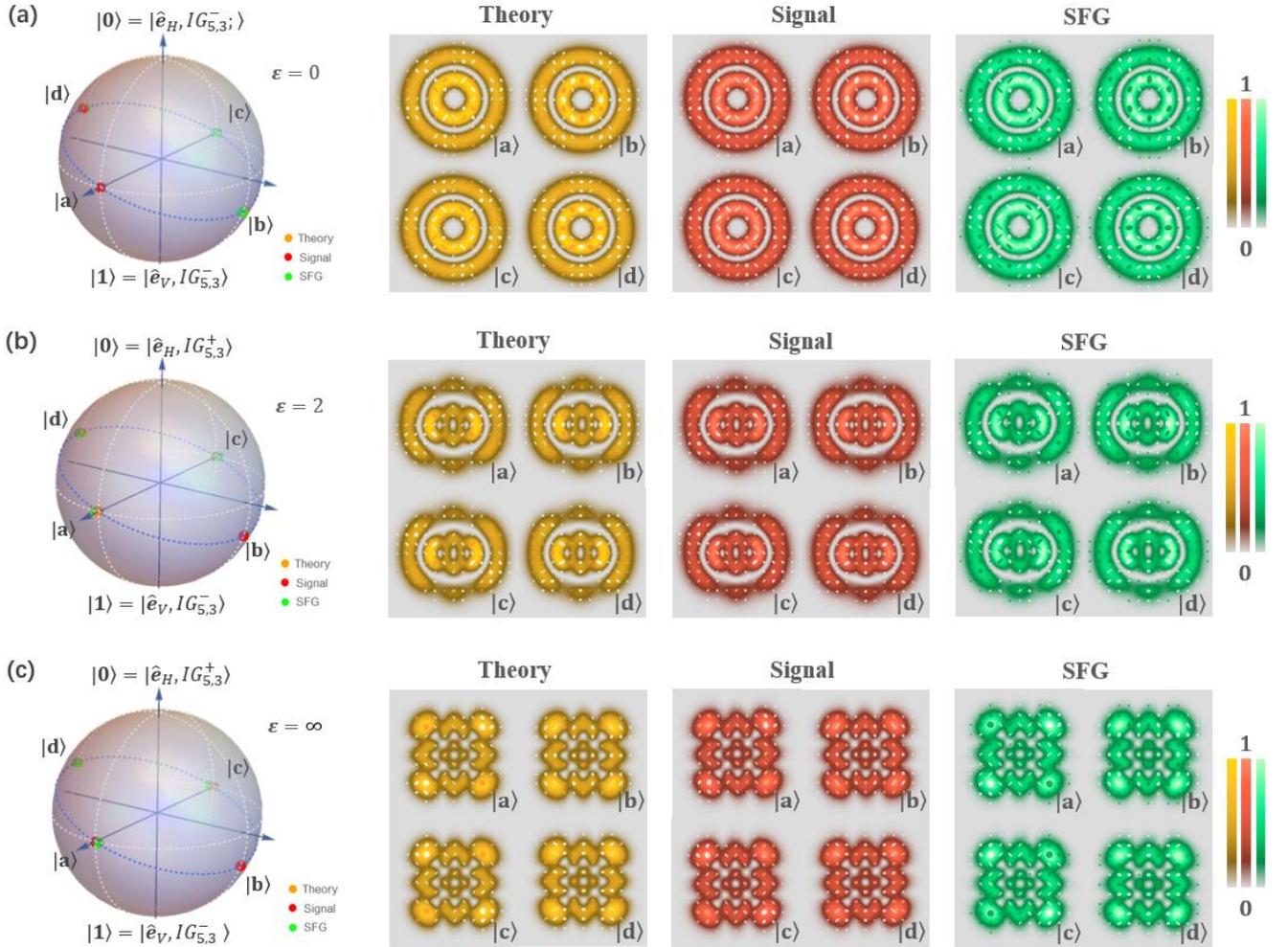

FIG. 4. Conformal upconversion for generalized SOC states defined in elliptical coordinates, where, in (a)–(c), the left sides are the theoretical and the measured SOC states on the HOPS, and the right sides are vector profiles of the corresponding states.

In the experiments, without loss of generality, $\hat{e}_{H/V}$ was first chosen as the SoP basis, and then we chose three groups, $IG_{53}^{\pm}$, with different ellipticities (i.e., $\varepsilon = 0$, 1, and $\infty$, respectively) as SoP-dependent modes. In case of $\varepsilon = 0$, $IG_{53}^{\pm}$ indeed became $LG_1^{\pm 3}$ that carried a well-defined OAM ($\pm 3\hbar$ per photon) with a radial structure $p=1$; for $\varepsilon = \infty$, $IG_{53}^{\pm}$ became a pair of helical HG modes, given by $HG_{53}^{\pm} = \sqrt{1/2}(HG_{32}^e \pm iHG_{23}^o)$. Note that for $\varepsilon > 0$, $IG_{Nm}^{\pm}$ can be represented as a superposition of LG modes of the same order $N = 2p + |\ell|$, given by [63]

$$IG_{Nm}^{+} = \sum_{p,\ell} c_{p,\ell} LG_p^{\ell}. \quad (6)$$

By using this familiar OAM eigen basis, we determined that the net OAMs carried by $IG_{53}^{\pm}(\varepsilon = 2)$ and $IG_{53}^{\pm}(\varepsilon = \infty)$ were $\pm 2.447\ \hbar$ and $\pm 3\ \hbar$ per photon, respectively (see Appendix A of SM for details). For each group, four states near the equator were chosen as the signals, as shown by the yellow points on the sphere and the associated vector profiles in Figs. 4(a)–(c). On the right side of the yellow patterns, the red and green vector profiles represent the experimentally observed signals and the associated upconversion, respectively, via spatial Stokes tomography. For all three groups, the transverse structures of the signals, in terms of both intensity and polarization, were in excellent agreement with their corresponding upconversions as well as their theoretical references, shown in yellow. Furthermore, on the HOPS, the positions of the experimentally prepared signals and the associated upconversions, shown by red and green points, respectively, were also highly proximate. All experimental results reflect the high fidelity of conformal upconversion for generalized SOC modes.

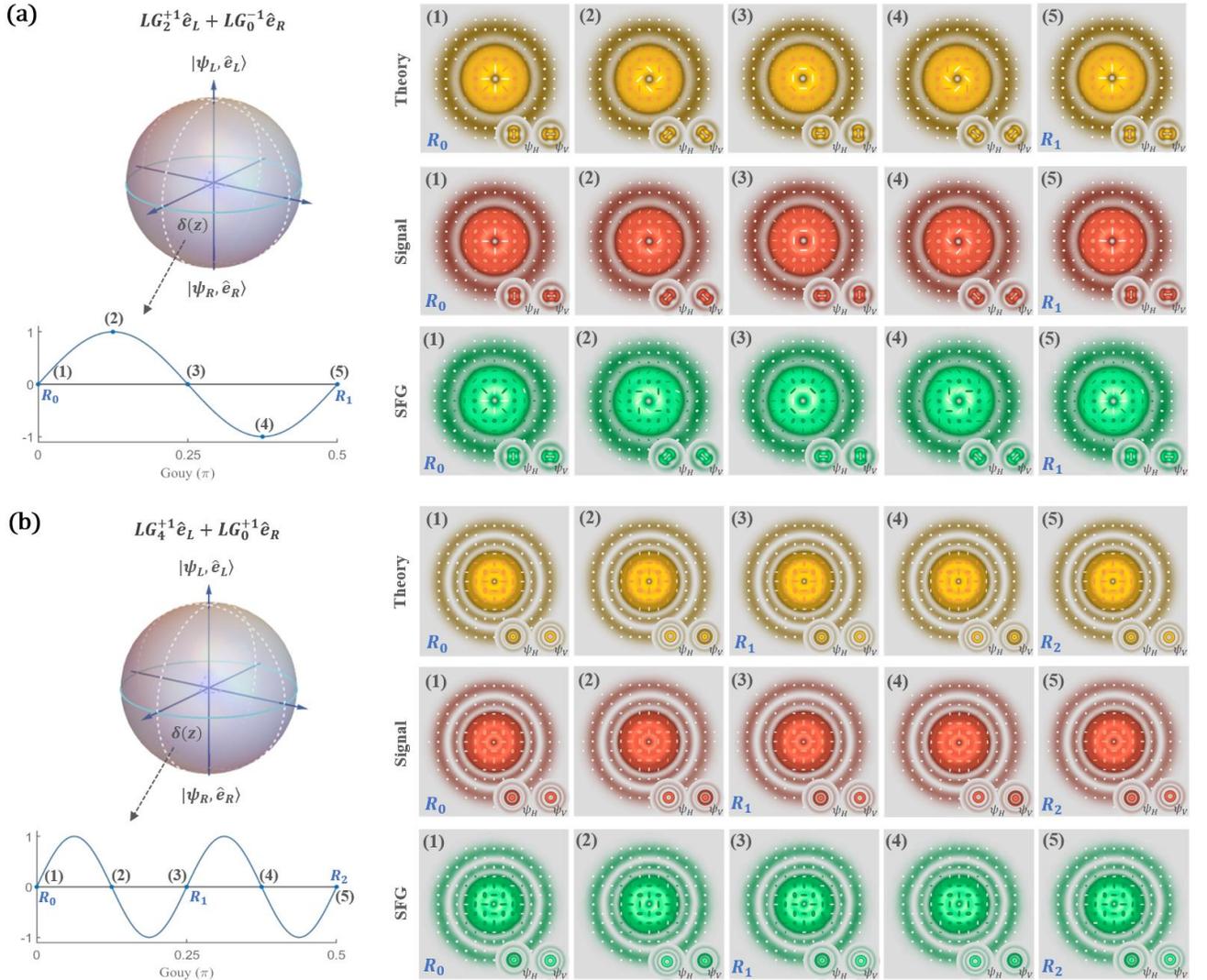

FIG. 5. Conformal upconversion of the vector Talbot mode. For both (a) and (b), the left-side schematics show the variation in the intramodal phase upon Gouy phase accumulation (0~0.5 pi); the patterns on the right side are the corresponding vector profiles in the specific propagation planes, denoted by (1)~(5).

## C. Conformal upconversion of non-eigenvector modes

Now, we further demonstrate more general cases in which the signals were non-eigen modes of the vectorial paraxial wave equation. The term "non-eigen" means that the spatial amplitude and polarization do not propagate in the same manner (wave vector), or rather, all components of the spatial-mode spectrum do not have an identical modal order, implying that the transverse structure would vary during propagation. Without loss of generality, we considered the simplest non-eigen vector mode containing only two modal orders, i.e., $N_1$ and $N_2$. The corresponding vector mode at a given propagation plane can then be expressed as

$$\sqrt{\alpha}\psi_{N_1}^{+}(\mathbf{r},z)\hat{e}_{+} + e^{i[\theta+\delta(z)]}\sqrt{1-\alpha}\psi_{N_2}^{-}(\mathbf{r},z)\hat{e}_{-}$$
$$\delta(z) = \Delta N \arctan(z/z_R) \quad . \quad (7)$$

Here, $\Delta N = N_1 - N_2$ can be either positive or negative integers that change the intramodal phase $\theta$ of the origin upon propagation via the term $\delta(z)$. As a consequence, the vector profile undergoes a continuous evolution during propagation to form a 3D polarization structure. Regarding frequency conversion for this 3D vector mode, the challenge is that the apparatus needs to be able to maintain all transverse structures of $E_{H/V}^{\omega_1}(\mathbf{r},z)$ and their relative relations in the region of light–matter interaction. Otherwise, any difference imposed during the conversion impacts the transverse structure and its subsequent propagation, i.e., the 3D polarization structure.

*Talbot modes*. — In the experiment, the non-eigen signals first examined were two vector Talbot modes, the vector profiles of which varied along the $z$-axis but periodically revived upon propagation, i.e., a property known as self-imaging, [51,64]. The two vector Talbot modes are given by

$$LG_2^{+1}(\mathbf{r},z)\hat{e}_L + e^{i[\theta+\delta(z)]}LG_0^{-1}(\mathbf{r},z)\hat{e}_R,$$
$$LG_0^{+1}(\mathbf{r},z)\hat{e}_L + e^{i[\theta+\delta(z)]}LG_4^{+1}(\mathbf{r},z)\hat{e}_R, \quad (8)$$

where $\Delta N$ is equal to four and eight for modes corresponding to $\delta(z_\infty) = 2\pi$ and $4\pi$, respectively. Accordingly, the original vector profile (denoted as $R_0$) of the first mode was revived only at the Fourier plane ($z_\infty$), denoted by $R_1$, and the structure of its 3D polarization was inversely symmetrical about the $z_R$ plane; the profile of the second mode had two revivals on the $z_R$ and $z_\infty$ planes, denoted by $R_1$ and $R_2$, respectively. Their theoretical structural evolutions are shown by the yellow patterns in Fig. 5. It is shown that the intensity profiles of both of them are propagation invariant but their polarization structure are not. Specifically, the propagation evolution of the first group manifests in a spatial-SoP rotation, i.e., a rotational CV mode, while the evolution of the second one exhibited the radial oscillation of the ellipticity. According to the theory, the two experimentally prepared and upconverted Talbot modes are represented by the red and the blue patterns in Fig. 5, respectively. We see that the 3D vector structures of the two signals were completely transferred into the upconversion with high precision.

*Optical Skyrmions*. — As a last example, we examined the case in which the vector profile of signal mode corresponds to 2D skyrmionic textures. Optical skyrmionic texture is an important property emerged in recent years that is extending the scope of the field of structured light; thus, it is worth to study the skyrmionic transformation in the conformal upconversion process. Skyrmions are a special type of quasiparticles with sophisticated topological textures [65], which have been studied in the realms of quantum fields, solid-state physics, and magnetic materials, but only recently it became topical in optical fields [66–68]. A typical example of optical skyrmions is the full Poincare vector beams, given by $LG_1^{+1}(\mathbf{r},z)\hat{e}_L + e^{i\theta}LG_1^0(\mathbf{r},z)\hat{e}_R$, where their Stokes vectors can construct skyrmionic textures in corresponding vector beams [69]. Theoretically, skyrmions can possess a myriad of topological states controlled by diverse topological numbers, polarity, vorticity, helical angle, to name a few [70]. The topological skyrmionic texture can be typically classified as Neel-type (hedgehog-like vector field) and Bloch-type (vortex-like vector field). Therefore, topological skyrmions can be used as effective toolkit to guide the topological state control of vector beams. Hereinafter, we demonstrate the skyrmionic vector beams will possess a topologically-invariant property in the SFG nonlinear conversion. Figure 6 shows the experimental results of skyrmionic textures before (red patterns) and after (green patterns) the nonlinear conversion, for different cases of Neel and Bloch types. The results clearly show that the spin textures were well maintained during the upconversion.

## IV. Conclusion

The conformal SFG based on type-0 QPM demonstrated above provides a high-efficient way to realize frequency conversion of arbitrary vector modes. This principle and its inverse process, i.e., down-conversion, can inspire many new applications involving vectorial structured light from both classical and quantum perspectives. Specifically, for applications such as optical communications, it can work as a frequency interface to link high-capacity (or -dimensional) classical (or quantum) channels carried by vector modes between different frequency ranges [59]. For laser-based applications [71,72], it extends current parametric laser technology to make it available to a diversity of vectorial structured beams/pluses, which is useful for many applications, such as optical tweezers, microscopy, and material processing. Moreover, beyond paraxial signals, this efficient and flexible (i.e., not limited to frequency doubling) conformal conversion paves the way to extend present upconversion imaging (or spectrum) technique into the polarization-resolved realm [3].

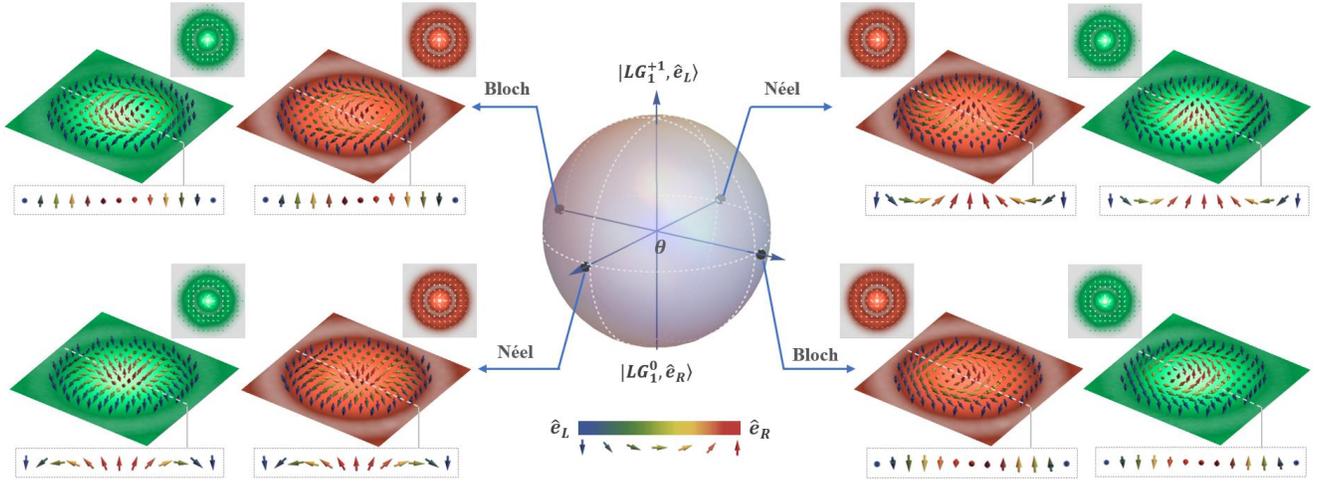

FIG. 6. Conformal upconversion for paraxial optical skyrmions defined by Stokes vectors of vector modes, where the upper (down) spin is defined by right- (left-) hand circular polarization.

After ensuring the fidelity of conversion, its efficiency is a focus in most applications. For information applications, i.e., the small-signal scenario demonstrated in the main text, the most effective way to enhance efficiency is by increasing the power of the pump. If a signal has a stable time sequence, such as that of pulse-pumped SPDC photons, a straightforward way to increase efficiency is to use a pulsed pump with the same sequence. For instance, the additional data in the Appendix B shown that one can easily arrive $\eta_q = 1$ for the small signal by using pulsed pump. If not, the use of a high-power continuous pump is unavoidable. In addition, note that in contrast to weak coherent light, the peak power of SPDC photons (usually ps-level pulses) used to determine the value of $\eta_q$ is much higher than their average power. For laser-based applications, first, the signal usually has a much higher power than that in the informational applications; second, $\eta_P$ in upconversion has a frequency-dependent gain, which was three for the SFG in this work (i.e., from 1560 nm to 520 nm). Thus, it is not difficult to achieve $\eta_P \geq 1$ by using type-0 QPM in the proposed setup. For instance, the additional data shown in the Appendix B of SM indicates that using a 500 mW pump is enough to fully convert a 200-mW pulsed infrared laser with vector $IG_{44}$ mode into the green. Finally, for all the cases, note that the beam size of pump should be customized to match the signal size. Namely, for low-order vector modes, using small-size pump can boost the efficiency significantly.

To summarize, we systematically examined the conformal frequency conversion based on non-degenerate SFG with type-0 QPM in this study. Our results show that by using the proposed methods and setup, a diversity of vector modes, including both 2D and 3D (i.e., eigen and non-eigen) modes, in the infrared region can be high-efficiently upconverted into those in visible regions without changing their full spatial structure. This proof-of-principle work provides a roadmap to build high-efficient frequency convertor for arbitrary vector-structured light/photons. Based on this demonstration, such novel applications as a high-dimensional quantum interface and polarization-resolved upconversion imaging will become feasible soon.

## ACKNOWLEDGMENT

This work was supported by the National Natural Science Foundation of China (Grant Nos. 62075050, 11934013, 61975047, and 12174185).

# Supplementary Materials for
# Conformal frequency conversion for arbitrary vectorial structured light

Hai-Jun Wu,[1] Bing-Shi Yu,[1] Zhi-Han Zhu,[1,*] Wei Gao,[1] Dong-Sheng Ding,[1,2] Zhi-Yuan Zhou,[1,2] Xiao-Peng Hu,[3] Carmelo Rosales-Guzmán,[1,4] Yijie Shen[5] and Bao-Sen Shi,[1,2,†]

[1] *Wang Da-Heng Center, Heilongjiang Key Laboratory of Quantum Control, Harbin University of Science and Technology, Harbin 150080, China*
[2] *CAS Key Laboratory of Quantum Information, University of Science and Technology of China, Hefei, 230026, China*
[3] *National Laboratory of Solid State Microstructures, College of Engineering and Applied Sciences, Nanjing University, Nanjing 210093, China*
[4] *Centro de Investigaciones en Óptica, A.C., Loma del Bosque 115, Colonia Lomas del campestre, 37150 León, Gto., Mexico*
[5] *Optoelectronics Research Centre, University of Southampton, Southampton SO17 1BJ, United Kingdom*

## Appendix A  Detailed Theoretical Frame

*Spatial Modes.* — In the simulation and data analysis, we used LG modes, denoted by $LG_p^{\pm\ell}$, and their superpositions to represent general spatial mode. The spatial complex amplitude of the LG mode in the cylindrical coordinates $\{r,\varphi,z\}$, with the spatial indices $\ell$ (azimuthal) and $p$ (radial), is given by [1]

$$LG_p^\ell(r,\varphi,z) = \sqrt{\frac{2p!}{\pi(p+|\ell|)!}}\frac{1}{w(z)}\left(\frac{\sqrt{2}r}{w(z)}\right)^{|\ell|}\exp\left(\frac{-r^2}{w_z^2}\right)\times L_p^{|\ell|}\left(\frac{2r^2}{w_z^2}\right)\exp\left[-i\left(kz+\frac{kr^2}{2R_z}+\ell\varphi-i\phi_g\right)\right], \quad (S1)$$

where $w_z = w_0\sqrt{1+(z/z_R)^2}$, $R_z = z^2 + z_R^2/z$, and $\phi_g = (2p+|\ell|+1)\arctan(z/z_R)$ denote the beam waist, radius of curvature, and Gouy phase upon propagation (here $z_R = kw_0^2/2$ is the Rayleigh length), respectively, and $L_p^{|\ell|}(\bullet)$ is the Laguerre polynomial with mode orders $p$ and $|\ell|$, given by $L_p^{|\ell|}(\gamma) = \sum_{k=0}^{p}[(|\ell|+p)!(-\gamma)^k]/[(|\ell|+k)!k!(p-k)!]$.

Ince-Gauss (IG) modes, used for representing the generalized SOC modes, are eigenfunctions of the paraxial wave equation in elliptical coordinates. A pair of IG modes with even and odd parity can be expressed as [2]

$$\begin{aligned}\text{IG}_{N,m}^e(\vec{r},z;\varepsilon) &= \gamma_e C_N^m(i\xi,\varepsilon)C_N^m(\eta,\varepsilon)\exp[G_N(r,z)]\\ \text{IG}_{N,m}^o(\vec{r},z;\varepsilon) &= \gamma_o S_N^m(i\xi,\varepsilon)S_N^m(\eta,\varepsilon)\exp[G_N(r,z)]\end{aligned}, \quad (S2)$$

where $\xi \in [0,\infty)$ and $\eta \in [0,2\pi)$ are the radial and the angular elliptic variables, respectively; $C_N^m(\bullet)$ ($S_N^m(\bullet)$) and $\gamma_e$ ($\gamma_o$) are even (odd) Ince polynomials and the associated normalization constants, respectively; and $G_N(r,z)$ is the amplitude envelop of Gaussian beams of order $N$. It should be noted that $\varepsilon \in [0,\infty)$ defines the ellipticity of the coordinates, i.e., the elliptical coordinates are not unique and, particularly, the IG modes become LG and HG modes with same the orders for $\varepsilon = 0$ and $\varepsilon \to \infty$, respectively. For a given $\varepsilon > 0$, $\text{IG}_{N,m}^{e(o)}$ can be expressed as a superposition of $(N+1)$ LG modes with the same order $N = 2p+|\ell|$, given by

$$\text{IG}_{N,m}^{e(o)} = \sum_p a_p \text{LG}_p^{\pm\ell}, \ p \in [0,1\ldots(N-1)/2], \quad (S3)$$

In addition, similar to the relation between the helical LG modes and their parity counterparts, IG mode can also exist in the helical manner, denoted as $IG_{Nm}^{\pm}$, that carry net OAM, and the corresponding relation is given by [3]

$$IG_{Nm}^{\pm} = \sqrt{1/2}(IG_{Nm}^e \pm iIG_{Nm}^o), \quad (S4)$$

Thus, $IG_{53}^{\pm}$ with $\varepsilon = 0$, 2, and $\infty$, used in the Sec. B of the main text, can be represented as LG superpositions modes by combined use of Eqs. (S1–S4), given by

---

$$IG_{5,3}^{\pm,\varepsilon=0} = LG_1^{\pm3}$$

$$IG_{5,3}^{+,\varepsilon=2} = 0.26LG_0^5 + 0.798LG_1^3 - 0.508\,LG_2^1$$
$$-0.165LG_2^{-1} - 0.103LG_1^{-3} - 0.005\,LG_1^{-5}$$

$$IG_{5,3}^{-,\varepsilon=2} = 0.26LG_0^5 + 0.798LG_1^3 - 0.508\,LG_2^1 \quad,\qquad\text{(S5)}$$
$$-0.165LG_2^{-1} - 0.103LG_1^{-3} - 0.005\,LG_1^{-5}$$

$$IG_{5,3}^{+,\varepsilon=\infty} = 0.791LG_0^5 - 0.5\,LG_2^1 - 0.354LG_1^{-3}$$

$$IG_{5,3}^{-,\varepsilon=\infty} = 0.791LG_0^{-5} - 0.5\,LG_2^{-1} - 0.354LG_1^{3}$$

from where we know they carry net OAM of $\pm 3\hbar$, $\pm 2.447\hbar$, and $\pm 3\hbar$ per photon, respectively.

*Modal Transformation.* — Here we derive the modal transformation of the general vector mode, shown in Eq. (1), in the Sagnac loop, i.e., $\mathbf{E}_S^{\omega 1}(\mathbf{r},z) = E_H^{\omega 1}(\mathbf{r},z)\hat{e}_H + e^{i\phi}E_V^{\omega 1}(\mathbf{r},z)\hat{e}_V$. By assuming the SoP relation $\hat{e}_+ = \sqrt{\beta}\hat{e}_H + e^{i\phi}\sqrt{1-\beta}\hat{e}_V$ and $\hat{e}_- = \sqrt{1-\beta}\hat{e}_H - e^{i\phi}\sqrt{\beta}\hat{e}_V$, we can obtain the SoP-dependent spatial modes with respective to $\hat{e}_H$ and $\hat{e}_V$, given by

$$E_H^{\omega 1}(\mathbf{r},z) = [\sqrt{\alpha\beta}u_+(\mathbf{r},z) + e^{i\theta}\sqrt{(1-\alpha)(1-\beta)}u_-(\mathbf{r},z)]e^{-ik(\omega 1)z}$$
$$E_V^{\omega 1}(\mathbf{r},z) = [\sqrt{(1-\alpha)\beta}u_+(\mathbf{r},z) - e^{i\theta}\sqrt{\alpha(1-\beta)}u_-(\mathbf{r},z)]e^{-ik(\omega 1)z} \qquad\text{(S6)}$$

Note that $E_H^{\omega 1}(\mathbf{r},z)$ and $E_V^{\omega 1}(\mathbf{r},z)$ are not usually orthogonal to each other, unless $\alpha = 0.5$ [4].

In addition, note that the premise to achieve the conformal upconversion, i.e., $E_H^{\omega 1}\hat{e}_H \to E_H^{\omega 3}\hat{e}_V$ and $E_V^{\omega 1}\hat{e}_V \to E_V^{\omega 3}\hat{e}_H$, is using flattop beam. Here we show briefly why the commonly used Gauss pump would lead to the angular spread of spatial spectrum, or rather, radial-modal degeneration for LG modes. The spatial amplitude of an SFG (with $k_3 = k_1 + k_2$) driven by two LG modes $LG_{p_1}^{\ell_1}(r,\varphi)$ with $k_1$ and $LG_{p_2}^{\ell_2}(r,\varphi)$ with $k_2$, can be expressed as

$$E_{SFG}(r,\varphi) = LG_{p_1}^{\ell_1}(r,\varphi)LG_{p_2}^{\ell_2}(r,\varphi)$$
$$= \frac{2}{\pi}\sqrt{\frac{p_1!p_2!}{(|\ell_1|+p_1)!(|\ell_2|+p_2)!}}\frac{(\sqrt{2}r)^{|\ell_1|+|\ell_2|}}{w_1^{|\ell_1|+|\ell_2|}w_2^{|\ell_1|+|\ell_2|}}\exp\left[-r\left(\frac{1}{w_1^2}+\frac{1}{w_2^2}\right)\right]\exp[-i(\ell_1+\ell_2)]L_{p_1}^{|\ell_1|}\left(\frac{2r^2}{w_z^2}\right)L_{p_2}^{|\ell_2|}\left(\frac{2r^2}{w_z^2}\right), \qquad\text{(S7)}$$

where $w_1$ and $w_2$ denote the beam waists of two pumps. Note that Eq. (S7) usually are no long a propagation-invariant (eigen) mode and one can further represent it in the form of a superposition of LG modes, i.e., the modal selection rule, given by

$$E_{SFG}(r,\varphi) = \sum_{\eta=0}^{n+j}a_\eta LG_\eta^m(r,\varphi), \qquad\text{(S8)}$$

where $m = \ell_1 + \ell_2$ and $n = (|\ell_1|+|\ell_2|-|\ell_1+\ell_2|)/2$, See Ref. 4 for more details. By assuming $w_1 = w_2$ and $\ell_2 = p_2 = 0$, we can obtain the modal transformation of LG modes in the upconversion pumped by a Gauss mode. For instance, the transformation of $LG_0^{+4}$, $LG_1^{+4}$, and $LG_2^{+4}$, i.e., a group data in Fig. 2(a), is shown below

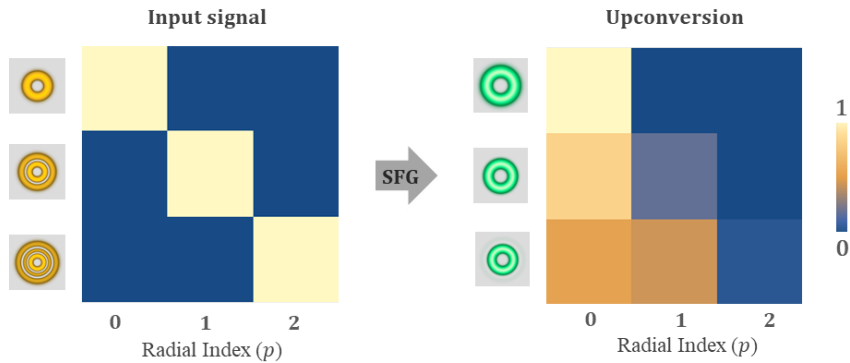

FIG. S1. Modal transformation of high-order LG modes during the SFG pump by Gauss-mode pump.
13

# Appendix B Additional Results

For the setup demonstrated in the main text, we also measured the efficiency of conversion in the depleted region by using pulse light, where the pulses were obtained via intensity-modulated 1560 nm wave and its SHG (780nm) wave. The quasi-continuous pulses have a 4ns duration with 100:1 duty factor. Before examine the conformal upconversion, the conversion efficiency between Gauss modes was characterized, as shown in Fig. S2, where the average power of signal was fixed at 1 mW. It was shown that, as illustrated in the Sec. A of the maintext, i) both $\eta_p$ and $\eta_q$ are linearly proportional to the pump power before the signal-depleted region; and ii) $\eta_q$ for 780 nm signal are double of that for 1560 nm, while $\eta_p$ is identical for the two. Besides, it is worthy to note that the upper limit of $\eta_q$ is below 100% in the Gauss-mode pumped SFG, even for a Gauss-mode signal.

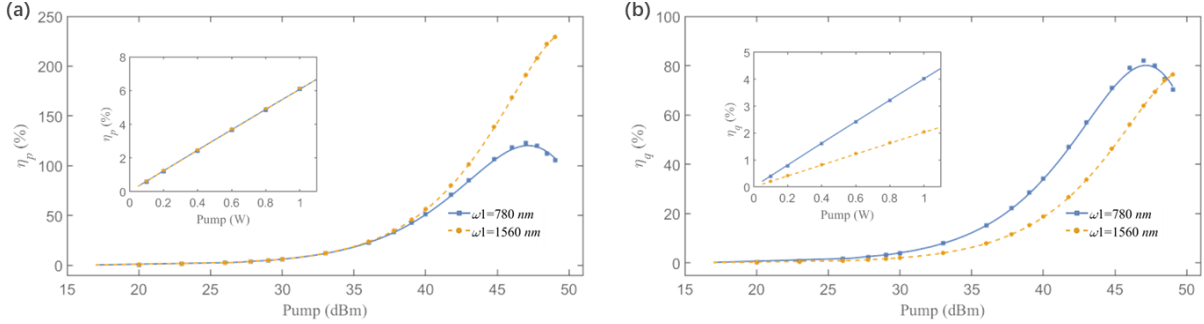

FIG. S2. Measured power (a) and quantum (b) efficiencies in the (1 mW) small signal SFG, where pump and signal were both Gauss modes with the same beam waist at the center plane of the crystal. The curves and dots are theoretical prediction and experimental data, respectively.

Then, we first consider the small signal in the depleted region. Specifically, a CV mode with $\ell=2$ at 780 nm was pumped by a flattop light at 1560 nm that can just well cover it, and the average power of the CV-mode signal was fixed at 1mW. The results in Fig. S3 show that, unlike the max $\eta_q$ was limited ~80% shown in Fig. S2, the 100% $\eta_q$ can be obtained experimentally by using only a 1.175 W pump with a super-Gauss mode. The vectorial transverse structure was also well maintained during the upconversion.

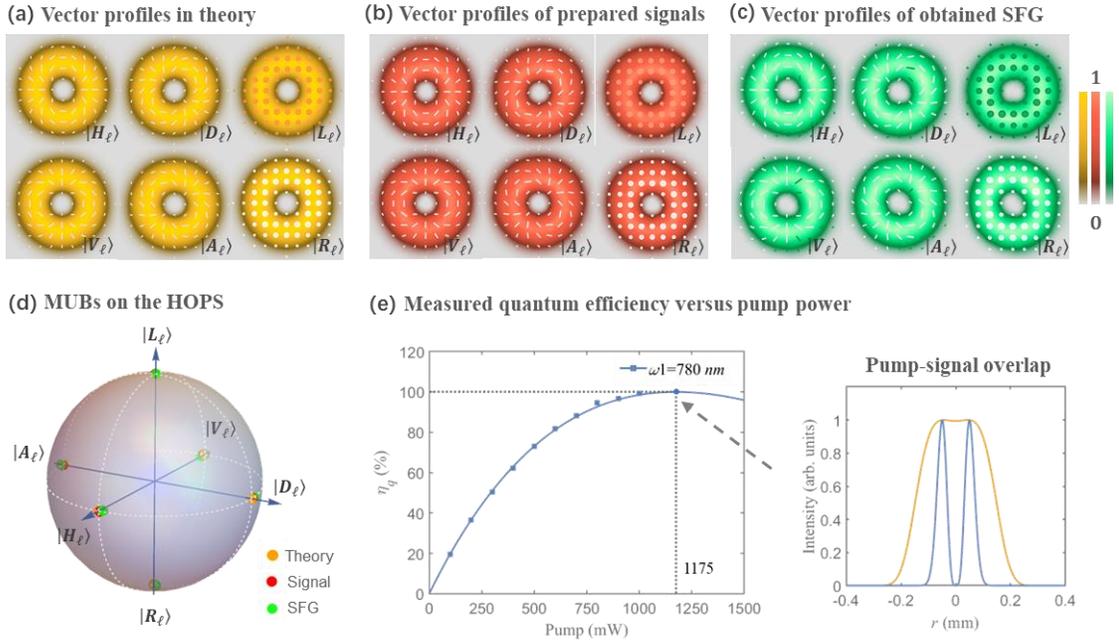

FIG. S3. Conformal upconversion of CV modes in the (1 mW) small-signal-depleted region, where the complete MUBs are defined in the SOC space spanned by $|L\rangle=|\hat{\mathbf{e}}_L,+2\rangle$ and $|R\rangle=|\hat{\mathbf{e}}_R,-2\rangle$. (a)–(c) show the theoretical vector profiles, the measured signals, and the corresponding SFG, respectively, at the 100% quantum efficiency, and corresponding positions of the MUBs on the HOPS are given in (d). (e) Measured quantum efficiency versus pump power.

Finally, we consider cases in laser-based applications, that is, depleted SFG with a larger signal. Here, a group of vectorial IG modes based on $IG_{44}$ with $\varepsilon=1$ were played as signals, whose power was 200 mW at 1560 nm. According to the measured $\eta_p$ with respect to power of pump shown in Fig. S4(a), one can obtain a 200 mW structured laser at 520 nm with the same vector profiles by using a 513 mW pump.

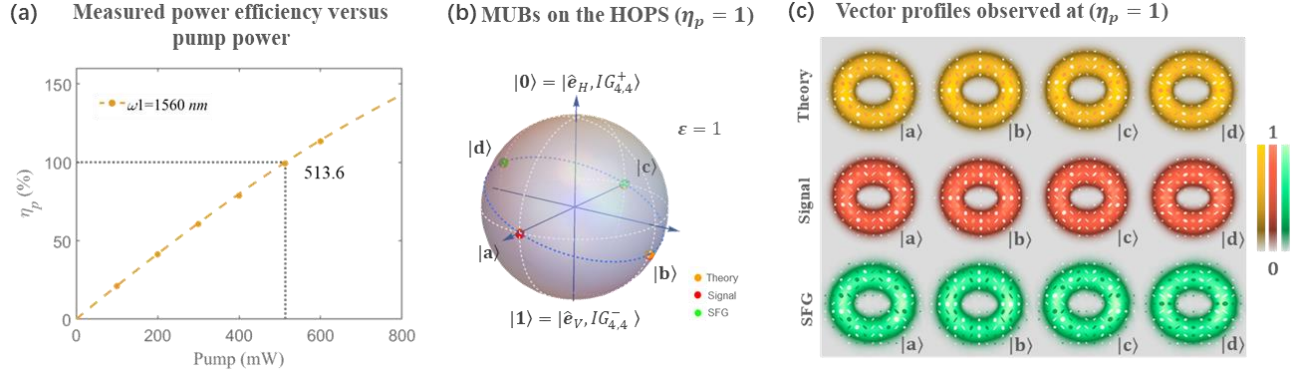

FIG. S4. Conformal upconversion of vectorial IG modes in the big-signal-depleted region, where the power of signal is 200 mw. (a) Measured power efficiency versus pump power. (b) and (c) show the SOC states on the sphere and associated vector profiles at the 100% power efficiency, respectively.